\PassOptionsToPackage{hypertexnames=false,bookmarks=false}{hyperref}
\documentclass[aps,pra,reprint,twocolumn,superscriptaddress,floatfix,nofootinbib,longbibliography]{revtex4-2}
\usepackage{graphicx,amsmath,amsfonts,amssymb,amsthm,xr}
\usepackage[T1]{fontenc}
\usepackage{graphicx}
\usepackage{mathrsfs}
\usepackage{float} 
\usepackage{booktabs}
\usepackage{makecell}
\usepackage{multirow}
\usepackage{indentfirst,csquotes}
\usepackage{soul}
\usepackage[acronym,indexonlyfirst,nomain,nohypertypes={acronym,notation}]{glossaries}
\usepackage{tikz}
\usepackage[colorlinks=true,linkcolor=blue,citecolor=blue,urlcolor=blue]{hyperref}
\usepackage{orcidlink}
\usepackage{amssymb,amsthm,amsmath}
\usepackage{xcolor,paralist,titlesec,fancyhdr,etoolbox}
\usepackage{lipsum}

\usetikzlibrary{quantikz2}

\makeatletter
\def\@evenhead{}
\def\@oddhead{}
\makeatother

\newacronym{QAOA}{QAOA}{Quantum Approximate Optimisation Algorithm}
\newacronym{QUBO}{QUBO}{Quadratic Unconstrained Binary Optimisation}
\newacronym{VQAs}{VQAs}{Variational Quantum Algorithms}
\newacronym{VQEs}{VQEs}{Variational Quantum Eigensolvers}
\newacronym{COBYLA}{COBYLA}{Constrained Optimisation by Linear Approximation}
\newacronym{CO}{CO}{Combinatorial Optimisation}
\newacronym{MPS}{MPS}{Matrix Product State}
\newacronym{SA}{SA}{Simulated Annealing}
\newacronym{QPU}{QPU}{Quantum Processing Unit}
\newacronym{CPU}{CPU}{Central Processing Unit}
\newacronym{ML}{ML}{Machine Learning}
\newacronym{SV-QAOA}{SV-QAOA}{Statevector QAOA}
\newacronym{MPS-QAOA}{MPS-QAOA}{Matrix Product State QAOA}
\newacronym{SV}{SV}{statevector}

\begin{document}
\title{Quantum Algorithm for Protein Side-Chain Optimisation: Comparing Quantum to Classical Methods}

\author{Anastasia Agathangelou \orcidlink{0009-0004-6056-9087}}
\thanks{Equal contributions, shared first-authorship.}
\affiliation{IBM Quantum, IBM Research Europe -- Zürich, 8803 Rüschlikon, Switzerland}

\author{Dilhan Manawadu\,\orcidlink{0000-0002-3575-8060}}
\thanks{Equal contributions, shared first-authorship.}
\affiliation{The Hartree Centre, STFC, Sci-Tech Daresbury, Warrington, WA4 4AD, United Kingdom}

\author{Ivano Tavernelli\,\orcidlink{0000-0001-5690-1981}}
\email{ita@zurich.ibm.com}
\affiliation{IBM Quantum, IBM Research Europe -- Zürich, 8803 Rüschlikon, Switzerland}

\date{\today}

\begin{abstract}
Modelling and predicting protein configurations is crucial for advancing drug discovery, enabling the design of treatments for life-threatening diseases. A critical aspect of this challenge is rotamer optimisation—the determination of optimal side-chain conformations given a fixed protein backbone. This problem, involving the internal degrees of freedom of amino acid side-chains, significantly influences the protein’s overall structure and function. In this work, we develop a resource-efficient optimisation algorithm to compute the ground state energy of protein structures, with a focus on side-chain configuration. We formulate the rotamer optimisation problem as a Quadratic Unconstrained Binary Optimisation problem and map it to an Ising model, enabling efficient quantum encoding. Building on this formulation, we propose a quantum algorithm based on the Quantum Approximate Optimisation Algorithm to explore the conformational space and identify low-energy configurations. To benchmark our approach, we conduct a classical study using custom-built libraries tailored for structural characterisation and energy optimisation. Our quantum method demonstrates a reduction in computational cost compared to classical simulated annealing techniques, offering a scalable and promising framework for protein structure optimisation in the quantum era.
\end{abstract} 

\maketitle

\let\thefootnote\relax

\section{Introduction}
Proteins are imperative to the drug discovery process, given that they are the building blocks of life~\cite{protein_func}. The ability to model and predict the correct spatial arrangement of all amino acids in a protein opens up incredible potential in the development of drugs against disorders in which they partake, as for instance in the treatment of certain forms of cancer and neurodegenerative diseases. Without a clear understanding of protein structure—and how even slight changes in composition can alter such structure—it is impossible to fully harness proteins for practical applications in drug discovery. To unlock their vast potential in this context, detailed and accurate structural studies are essential. Such investigations are imperative and must be conducted with the highest level of rigour and precision. 

Proteins are comprised of a specific sequence of amino acids (also called residues), each of which is characterised by a different side-chain. The exact sequence of amino acids fully determines the way the amino acid chain folds, thereby giving rise to the 3-dimensional structure of the protein~\cite{PMC2443096}. Since the function of a protein is intimately related to its structure~\cite{protein_britannica}, considerable effort has been made to understand protein folding. In addition to main-chain folding, optimising side-chain conformations—specifically the internal degrees of freedom known as rotamers—is essential for accurate protein modelling. Rotamers represent discrete conformational states defined by torsional angles around side-chain bonds. Efficient exploration and optimisation of this combinatorial space are critical for reliable protein structure prediction, stability assessment, and rational drug design.

Protein structure prediction can be approached numerically via several methods. One common approach involves statistical mechanics, which samples the protein’s configurational space through techniques such as Monte Carlo simulations. However, this method faces the challenge that the configurational space grows exponentially with the number of residues, making exhaustive sampling computationally expensive. Another widely used technique is molecular dynamics, which solves the equations of motion based on predefined Molecular Mechanics force fields \cite{MM} to calculate forces and energies. The main limitations here are the accuracy of the force fields and the relatively short timescales (in the order of $\mu s $ to $ms$) accessible by simulations, typically not exceeding milliseconds. More recently, \gls{ML} methods have been applied, either alone or in combination with traditional approaches. While \gls{ML} can greatly accelerate predictions, the reliability of these methods depends heavily on the completeness of the training data, and they may struggle to accurately predict novel or rare protein folds that are not included in the training set. In recent years, classical methods have made significant progress in protein structure prediction, most notably with the advent of AlphaFold~\cite{AlphaFold, Alphafold-2}. Although such models have achieved high accuracy on many known proteins, they are primarily trained on experimentally determined structures and may have limitations when applied to biomolecular systems that differ significantly from the training data.

Solving classically hard computational tasks has long been a driving motivation for quantum computing. Several quantum algorithms have been proposed to address the protein folding problem and configuration optimisation. The motivation stems from the inherent difficulty of the problem, which is NP-complete even in its minimal formulation on a lattice, as shown in \cite{NP-hard-PF}. When extended beyond the lattice model to incorporate side-chain conformations \cite{10.1007/11415770_32, https://doi.org/10.1002/1096-987X(200008)21:11<999::AID-JCC9>3.0.CO;2-A}, the problem retains its NP-completeness. Quantum algorithms have become a promising approach to this problem, particularly through the use of protein lattice models \cite{Levitt, Hinds}. These models significantly reduce the conformational search space, thereby mitigating the computational cost associated with exhaustive sampling. Babbush et al. \cite{Babbush_2014, Perdomo_2008} set the foundations for mapping the protein folding problem into a structure compatible with adiabatic quantum computation, employing an adiabatic quantum evolution process to evolve an initial state to a final state which encodes the minimum energy state. Moreover, quantum annealing—a heuristic implementation of adiabatic quantum computing which can be implemented in a hardware-oriented fashion for near-term devices with limited coherence and connectivity—has been applied to discrete lattice models \cite{cai2014practicalheuristicfindinggraph, Irb_ck_2024, panizza2024proteindesignintegratingmachine}. These models, characterised by local interactions, align more naturally with the sparse connectivity and interaction constraints of quantum annealing hardware. However, these approaches—and others in the same line \cite{babej2018coarsegrainedlatticeproteinfolding, perdomoortiz2012findinglowenergyconformationslattice}—still incur substantial resource costs due to constraint enforcement and hardware limitations, thus fundamentally limiting scalability. As focus shifted towards gate-based quantum computing, new approaches emerged to adapt protein folding problems to universal quantum architectures. Several approaches in the coarse-grained framework of this problem have also raised concerns about the scalability and effectiveness of the variational algorithm, the \gls{QAOA}, citing challenges such as high resource requirements and difficulties in constraint enforcement \cite{boulebnane2022peptideconformationalsamplingusing, Linn_2024}. Mustafa et al. \cite{mustafa2022variationalquantumalgorithmschemical} compared \gls{QAOA} and \gls{VQEs}, finding that \gls{VQEs} often produced better energy estimates in their benchmarks, but also encountered unphysical results and significant difficulty in enforcing constraints via penalty terms. To improve optimisation efficiency in \gls{QAOA}, Fingerhurth et al. \cite{fingerhuth2018quantum} tackled the lattice protein folding problem using the quantum alternating operator ansatz. Their approach introduced novel XY and XZ mixer Hamiltonians designed to penalise short- and long-range overlaps in coarse-grained lattice models, thereby enhancing optimisation efficiency. More recently, universal gate-based methods have continued to evolve. Pamidimukkala et al. \cite{Pamidimukkala_2024} explored broader quantum circuit architectures for protein modelling, and Romero et al. \cite{romero2025proteinfoldingalltoalltrappedion} proposed encoding higher-order unconstrained binary optimisation problems on all-to-all connected trapped-ion quantum processors, paving the way for more complex folding models. In parallel, hybrid quantum-classical optimisation techniques for improving resource requirements have emerged. Robert et al. \cite{Robert_2021} proposed a resource-efficient approach that combines a VQE with a classical genetic algorithm to tune the quantum circuit parameters and optimise coarse-grained protein folding configurations. This method can capture the secondary and tertiary protein structure at a coarse-grained level by modelling backbone and side-chain interactions on a tetrahedral lattice. However, the reconstruction of side-chains with the internal rotational angles remains an open challenge for achieving full biological relevance in protein folding.

Side-chains play a crucial role in the protein folding problem \cite{side-chain-cluster}, and so various approaches have also been developed to address the side-chain optimisation problem. One notable development involves integrating Rosetta, a widely-used protein design software, with D-Wave’s quantum annealing hardware~\cite{Mulligan752485}. The resulting \textit{QPacker} algorithm reformulates the side-chain packing problem into a \gls{QUBO} form, which is solved via quantum annealers such as the D-Wave 2000Q system, supplemented by classical post-processing with \textit{qbsolv} \cite{qbsolv}. While QPacker is based on a discrete rotamer library, more recent work has explored continuous representations. Casares et al. \cite{Casares_2022} developed QFold, which uses a neural network to predict likely torsion angles of the side-chains and then implements a coined quantum walk to update the torsion angles of both the backbone and the side-chains with a Metropolis acceptance criterion. This method models torsion angles as continuous variables rather than discrete sets, enabling finer structural granularity but also expanding the search space significantly. 

In this work, we consider a fixed protein backbone (fold) and focus on the challenging problem of optimising the internal degrees of freedom of the side-chains, collectively known as rotamers, to minimise the energy of the protein. This represents the next step beyond coarse-grained modelling on the path to full protein folding on a quantum computer. Building upon the quantum-enhanced methodologies discussed previously, we develop a quantum algorithm for side-chain optimisation by implementing \gls{QAOA} in combination with a local XY mixer to solve the corresponding \gls{QUBO} problem. This paper is organised as follows: in Section \ref{sec:pipeline}, we propose a complete framework for the protein folding side-chain optimisation problem in Qiskit\cite{qiskit2024}. We will show a detailed pipeline that begins with a linear chain of amino acids, including all of their respective side-chains. This is followed by the modelling and encoding of this structure, and finally the optimisation of such using \gls{QAOA}. In Section \ref{sec:computational_details} we discuss the computational details of the \gls{QAOA} algorithm, giving the numerical details of the experiments conducted. Finally, Section \ref{sec:results} reviews the results. We outline the possible constraint-imposing techniques and the associated resource requirements for the problem. Then, we discuss the benefits of our quantum algorithm for protein side-chain optimisation over classical approaches and perform a comparative analysis of the scaling behaviour between the two methods. In particular, we will examine how the heuristic quantum method scales relative to an established heuristic classical method—\gls{SA}—and give an estimate of a crossing point, i.e. the point at which the quantum method may outperform the classical~\cite{Abbas_2024, Baiardi_2023}.
\begin{figure}[b]
    \centering
    \includegraphics[width=\linewidth]{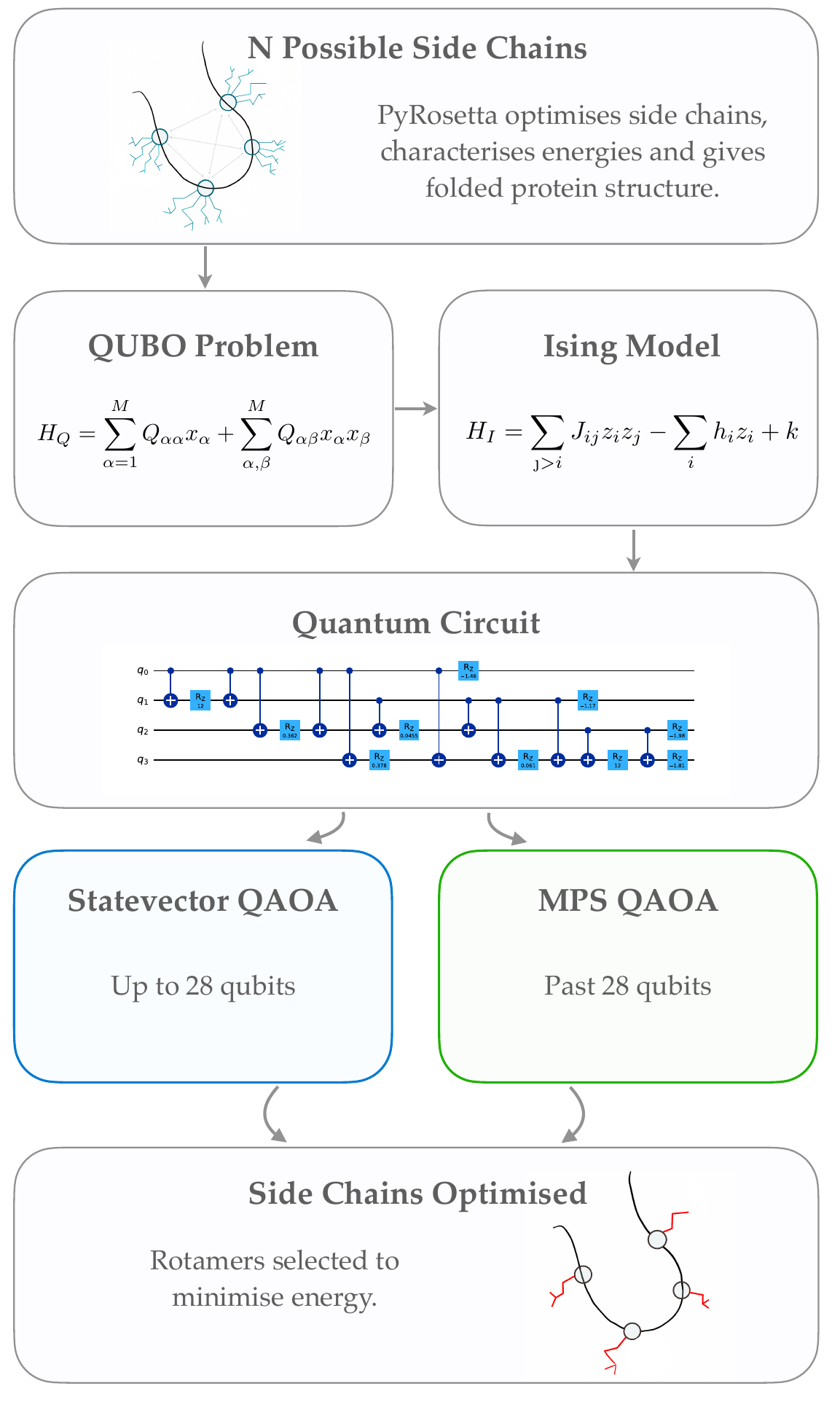}
    \caption{Workflow of the paper: from protein backbone folding, through model mapping, to quantum optimisation for identifying the ground state and optimal side-chain configurations.}
    \label{fig:workflow}
\end{figure}

\section{Protein side-chain optimisation: a hybrid quantum-classical  workflow}
\label{sec:pipeline}
Solving the protein folding problem involves finding the minimum energy geometry of a protein given a sequence of amino acids as input. 
This problem can be approached at multiple levels of resolution, from coarse-grained models (amino acid) to atomic. 
While coarse-grained models are typically sufficient to capture the global fold of a protein, atomic-resolution models are essential for refining finer structural details, such as side-chain conformations. These details become particularly important when examining protein–substrate or protein–drug interactions, where the presence of an external ligand can induce changes in side-chain orientations that must be re-optimised. The high specificity of such intermolecular interactions makes \gls{ML} techniques based purely on training data less well-suited for this task.

In this work, we focus specifically on side-chain optimisation, assuming a fixed backbone structure and seeking the lowest energy. We choose to model it as a \gls{CO} problem. With a protein consisting of $N$ residues, each of which has $n$ rotamers, there are $n^N$ possible configurations. Finding the exact solution to this NP-hard problem therefore scales exponentially. In this work, we optimise the coarse-grained structure classically but use a quantum algorithm to determine the choice of rotamers. We classically determine a table of all the nearest-neighbour pairwise interaction energies of rotamers. Then, given this table of energies, the optimisation problem is to choose a set of rotamers that minimises the energy. The problem is formulated as a \gls{QUBO}, which can then be translated into a qubit-based representation. A solution is then found by means of the \gls{QAOA} algorithm. The pipeline presented in this section starts with the preparation of protein structures, including classical optimisation, and proceeds through problem formulation as an Ising model and its encoding onto qubits for quantum optimisation. The goal is to identify the ground-state energy corresponding to selected rotamer configurations, using two parallel quantum approaches—\gls{SV-QAOA} and \gls{MPS-QAOA}, as illustrated in the workflow shown in Fig.~\ref{fig:workflow}. 

\subsection{Protein Structure Preparation}
\begin{figure}[t]
    \centering
    \includegraphics[width=0.85\linewidth]{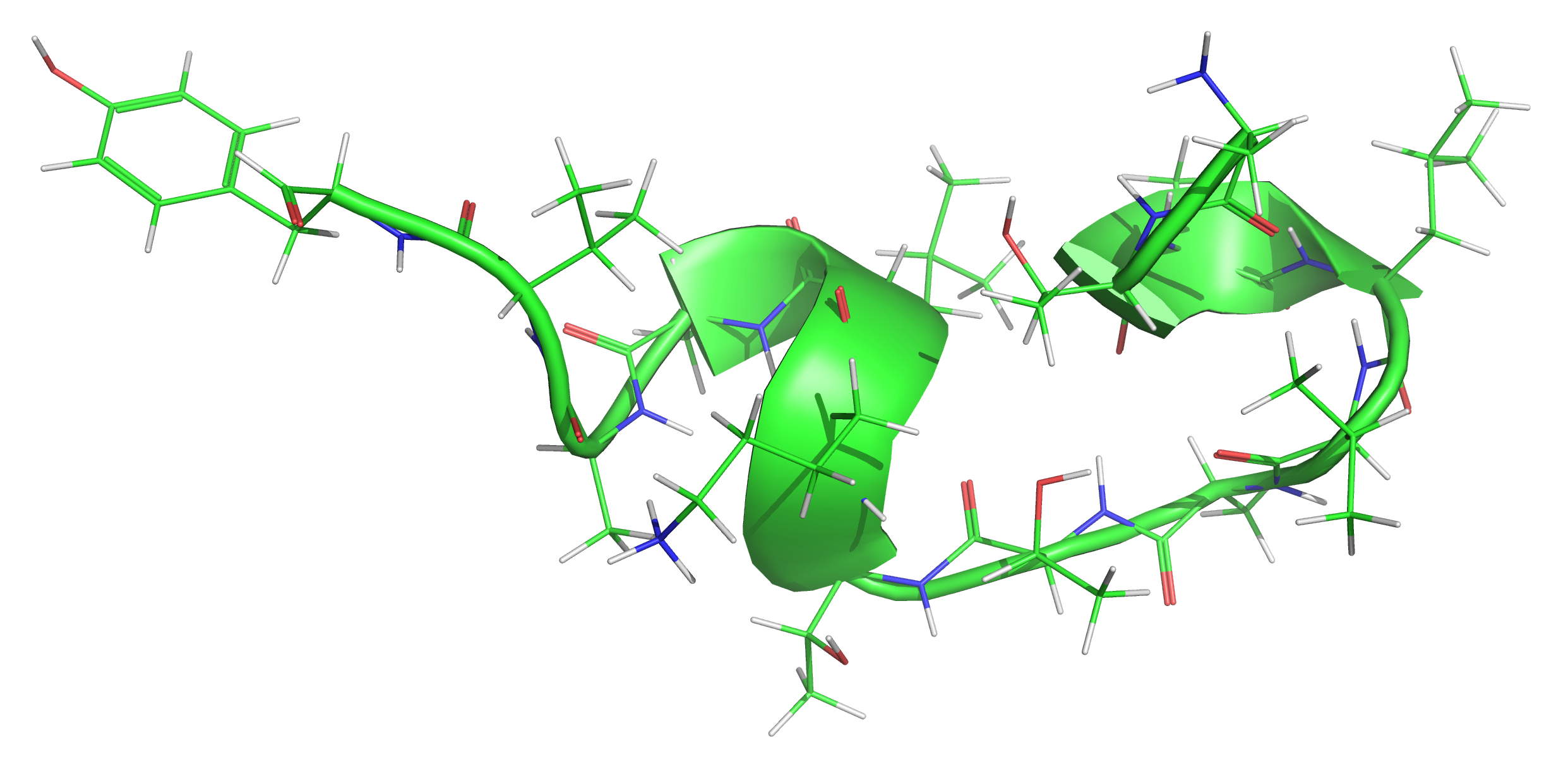}
    \caption{Visualisation of a 14-residue protein structure in PyMOL in the $x-y$ plane. The protein backbone is depicted in green, nitrogens atoms in blue, oxygen atoms in red, and hydrogen atoms in white.}
    \label{fig:pymol}
\end{figure}

The protein structures used in this workflow are acquired and prepared in a generic way, permitting reproducibility and a wide scope of test structures. A PDB (Protein Data Bank) file, which describes the three-dimensional structure of a molecule, is obtained from the Protein Data Bank~\cite{PDB}. Through PEPstrMOD, the tertiary structure of small peptides with sequence lengths between 7 and 25 residues can be predicted, generating a PDB file based on structural templates derived from X-ray crystallography~\cite{pepstrmod}. The structure is then protonated with PyMOL~\cite{PyMOL} and can be visualised as in Fig. \ref{fig:pymol}. Strictly, one should consider a minimum of seven amino acids, as this is approximately the length at which folding effects begin to emerge~\cite{https://doi.org/10.1002/jcc.20543}.
Where testing on smaller systems here, we truncate a given 7 amino acid polypeptide to the desired size. 

Preliminary evaluation of the interaction energies among all possible rotamer pairs indicates that non-nearest-neighbour contributions are negligible, contributing minimally to the system’s total energy—see Appendix~\ref{app:rot_int}. Therefore, in this work, we consider only the interactions between rotamers of nearest-neighbour amino acids. While this restriction reduces the number of terms in the Hamiltonian and improves computational efficiency in practice, the problem remains NP-hard in general. Long-range interactions can however be reintroduced in subsequent analyses to obtain a more complete and accurate structural evaluation, if required, without significant additional complexity in the calculations.

\subsection{Two-body energy cost function using classical pre-processing}
\begin{figure}[b]
    \centering
    \includegraphics[width=0.72\linewidth]{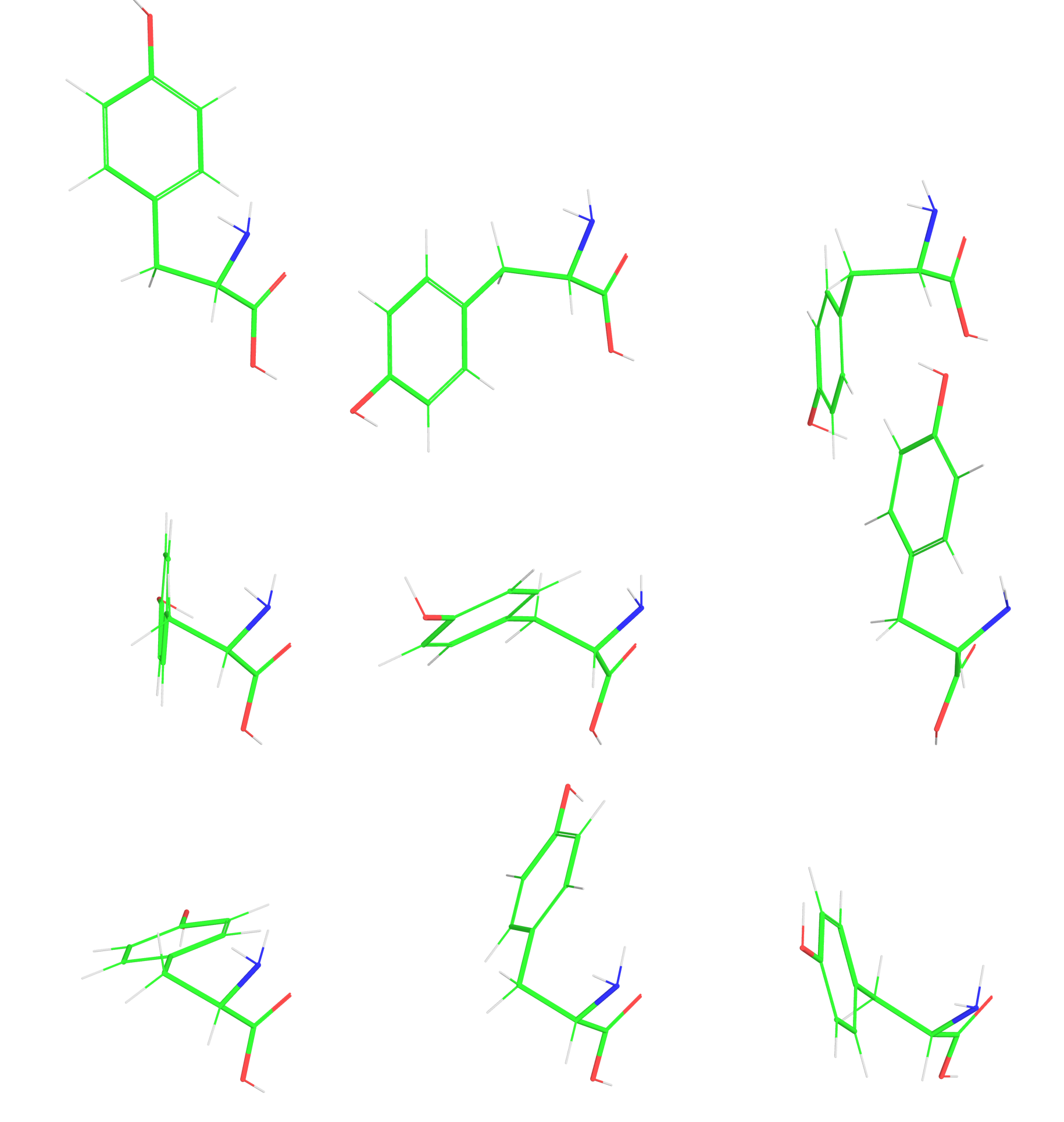}
    \caption{Visualisation of the rotamer set for the amino acid Tyrosine (TYR). The figure displays nine distinct side-chain conformations on a single residue, generated by rotations about the rotatable $\chi$ dihedral angles. Backbone atoms are shown in green, nitrogen atoms in blue, oxygens atoms in red, and hydrogen atoms in white.}
    \label{fig:TRP_rotamers}
\end{figure}

The next step in the workflow consists of the generation of the side-chain conformers and the evaluation of all possible interaction energies between consecutive amino acids along the protein sequence. Here, we make use of PyRosetta~\cite{10.1093/bioinformatics/btq007}, which is a toolkit for computational modelling and analysis of protein structures. PyRosetta includes a built-in energy evaluation function (score function) that assigns energy terms, such as electrostatics and van der Waals, to the different configurations. This scoring function is used to guide structure refinement protocols of the protein towards initial candidate low-energy conformations by iteratively alternating between ``packing''—a procedure that selects side-chain rotamers to reduce steric clashes and provide an initial screening of viable conformations—and local minimisation of atomic coordinates to further reduce the energy. The result of this process is a PDB file of the refined protein, including all possible rotamers identified during the initial screening (a set of rotamers for a given residue can be observed in Fig. \ref{fig:TRP_rotamers}), along with a table of one- and two-body interaction energies associated with them. These energies serve as the inputs to the \gls{CO} problem, i.e. we will search this table of energies to identify the combination of rotamers that minimises the overall energy, yielding the final ground-state conformation of the protein.

\subsection{Mapping to a \gls{QUBO} Hamiltonian}
With the one-body energy terms and pairwise rotamer interaction energies obtained, we proceed to construct the Hamiltonian. The objective matrix to be optimised is based on an Ising model and defines a classical Hamiltonian that models all pairwise interactions between the rotamers across neighbouring residues. To this end, we first formulate the problem as a \gls{QUBO} model \cite{Gilliam_2021}, which we then convert into an equivalent Ising Hamiltonian. This Hamiltonian is subsequently mapped onto a qubit-based representation suitable for quantum computation. Note that our model is defined by quadratic terms and will be characterised by a sum of the two-body interaction terms. Extensions to higher order terms are possible, but not considered in this work.

\subsubsection{Model Derivation}
In this section, we derive the QUBO representation of the side-chain interaction Hamiltonian. Residues are labelled $R_i$ for $i \in [1,N]$ and the rotamers of $R_i$ are the set $\{ p^i_{\alpha}\}_{\alpha=1,n}$.
For each rotamer there is a self (one-body) energy term $E_{self}(p^i_{\alpha})$ and for every pair of rotamers there is an interaction (two-body) energy term, $E_{int}(p^i_{\alpha},p^j_{\beta})$. A configuration of rotamers is given by the vector $\vec{p} = (p^1_{\alpha},p^2_{\beta},...,p_{\nu}^N)$, and we can write the total energy (cost) of this configuration as,
\begin{equation}
        E(\vec{p}) = \sum_{i=1}^N E_{self}(p^i_{\alpha}) + \sum_{i<j}^N E_{int}(p^i_{\alpha}, p^j_{\beta}).
\label{eq:Total-energy}
\end{equation}

At this point, we can map the problem to binary decision variables $x_{i} \in \{0,1\}$, where each variable corresponds to a specific rotamer. For a given residue with $n$ possible rotamers, we introduce $n$ decision variables $\{x_i\}_{i=1,n}$ that record which of the rotamers is chosen. For each residue, the choice of rotamer is a length-$n$, weight-1 bitstring $x_0 x_1 x_2 ... x_{n-1}$, where exactly one bit is set to 1, indicating the selected rotamer. For a system with $N$ residues, the full configuration is represented by a length-$Nn$, weight-$N$ bitstring $x_0 x_1 x_2 ... x_{Nn-1}$, partitioned into $N$ blocks of size $n$. Each block corresponds to a residue and contains exactly one bit set to 1. The weight constraints arise from the physical condition that each residue takes exactly one rotational conformation.

Based on the table of self and interaction energies, we construct a symmetric matrix $Q$ such that the problem of minimising Equation \eqref{eq:Total-energy} reduces to finding the bitstring $x^* = \min_x x^T Q x$. The entries of the matrix $Q$ encode the energetic landscape of all possible rotamer configurations. The diagonal elements, $Q_{\alpha\alpha}$, represent the self-energies of individual rotamers (i.e., one-body terms), while the off-diagonal elements, $Q_{\alpha\beta}$, capture the pairwise interaction energies between rotamers $\alpha$ and $\beta$ on different residues. We can write the cost function $C$ more explicitly as,
\begin{equation}
    C = \sum_{\alpha=1}^{M} Q_{\alpha\alpha} x_\alpha + \sum_{\alpha,\beta}^M Q_{\alpha\beta} x_{\alpha}x_{\beta},
\label{eq:qubo-full}
\end{equation}
where we define the total number of rotamers $M = Nn$. As an illustrative example, we consider the simple case of a protein composed of 4 residues $\{R_1,...,R_4\}$, each with 2 possible rotameric states (configurations) labelled $\{A,B,\dots, H\}$, as shown in Table~\ref{tab:possible_rots}. Equation \eqref{eq:Q_matrix} presents an explicit example of the corresponding $Q$ matrix. The indices $\alpha, \beta \in \{A,B,\dots,H\}$ label individual rotamers as listed in Table~\ref{tab:possible_rots}, with one-body energy terms on the diagonal and two-body interaction terms appearing in off-diagonal blocks. Since only the interactions between rotamers on neighbouring residues are considered, the matrix $Q$ has a block-banded structure and compactly encodes all energetic information for the system. 
\begin{table}[ht]
    \centering
    \small
    \setlength\tabcolsep{6pt}  
    \renewcommand{\arraystretch}{1.2}  
    \begin{tabular}{|c|c|c|c|c|}
        \hline
        {Config.} & ${R_1}$ & ${R_2}$ & ${R_3}$ & ${R_4}$ \\
        \hline \hline
        1 & A & C & E & G \\
        \hline
        2 & B & D & F & H \\
        \hline
    \end{tabular}
    \caption{Two possible rotamer configurations (A, B, C, ...) of a 4-residue peptide ($R_1, R_2, R_3, R_4$).}
    \label{tab:possible_rots}
\end{table}
{\small{
\begin{equation}
    Q = 
    \begin{bmatrix}
    Q_{AA}&0&Q_{AC}&Q_{AD}&0&0&0&0\\
    0&Q_{BB}&Q_{BC}&Q_{BD}&0&0&0&0\\
    Q_{CA}&Q_{CB}&Q_{CC}&0&Q_{CE}&Q_{CF}&0&0\\
    Q_{DA}&Q_{DB}&0&Q_{DD}&Q_{DE}&Q_{DF}&0&0\\
    0&0&Q_{EC}&Q_{ED}&Q_{EE}&0&Q_{EG}&Q_{EH}\\
    0&0&Q_{FC}&Q_{FD}&0&Q_{FF}&E_{FG}&Q_{FH}\\
    0&0&0&0&Q_{GE}&Q_{GF}&Q_{GG}&0\\
    0&0&0&0&Q_{HE}&Q_{HF}&0&Q_{HH}
    \end{bmatrix} 
    \label{eq:Q_matrix}
\end{equation}
}}
While physically only a single rotamer can be chosen for each residue, solving the unconstrained optimisation problem in Equation \eqref{eq:qubo-full} may yield solutions that violate this constraint, resulting in unphysical solutions. As such, this condition needs to be imposed as a set of constraints. As mentioned above, the full length-$M$ bitstring can be decomposed into $N$ length-$n$ substrings that each correspond to one of the $N$ residues. The requirement that exactly 1 rotamer is chosen translates to requiring that each substring has a Hamming weight of 1, and therefore the full bitstring has a Hamming weight of $N$. For example, in the case shown in Table~\ref{tab:possible_rots}, each substring is length 2 and the physically allowed substrings are $01$ and $10$.

There are several ways to improve the probability of only sampling allowed states. One of the main approaches is to implement so-called soft constraints. This involves adding penalty terms to the cost function that strongly discourage physically disallowed states. In general, we can either enforce the Hamming weight globally by discouraging any bitstring with Hamming weight other than $N$, or we can impose local constraints that discourage substrings of Hamming weight not equal to 1. We opt for local penalty terms as bitstrings corresponding to physical systems must conserve the Hamming weight locally. Specifically, we add two-body Pauli $ZZ$ terms between each pair of rotamer qubits within the same residue, scaled by a positive coefficient. These terms discourage any bitstring that encodes multiple active rotamers within the same residue block. Note that it is also possible to implement so-called hard constraints that do not involve modifying the Hamiltonian. We discuss a hard constraint method in Section \ref{sec:sidechain_qaoa} and compare the benefits of the local penalty and hard constraint method in Section \ref{sec:circuit_depth}.

\subsubsection{Mapping to the Ising Hamiltonian}
To map the \gls{QUBO} problem to a qubit-based formulation, we first convert to an Ising model with a simple change of variables $z_{i} = 1 - 2x_{i}$. Performing this substitution and defining $J = 0.25Q$ and a vector $h_{i} =1/2 Q_{ii} + \sum_{j \geq i} 1/4 Q_{ij}$, gives the Ising form
\begin{equation}
    H_I = \sum_{j>i} J_{ij}z_{i}z_{j} - \sum_{i} h_{i}z_{i} + k
\label{eq:Ising-Ham}
\end{equation}
where $k = \sum_{i} Q_{ii}/2 + 1/2 \sum_{i \neq j} Q_{ij}/4$ is a constant term to be added back on after the optimisation. The final step is to map the Ising Hamiltonian in Equation \eqref{eq:Ising-Ham} to a qubit Hamiltonian by promoting spin variables $z_i$ to Pauli operators as follows,
\begin{equation}
    z_{i} \mapsto I_{0} \otimes I_{1} \otimes ... \otimes I_{i-1} \otimes Z_{i} \otimes I_{i+1} \otimes ... \otimes I_{n}.
\end{equation}

\subsection{Side-chain optimisation with \gls{QAOA}}
\label{sec:sidechain_qaoa}
\begin{figure}[H]
     \centering
     \includegraphics[width=\linewidth]{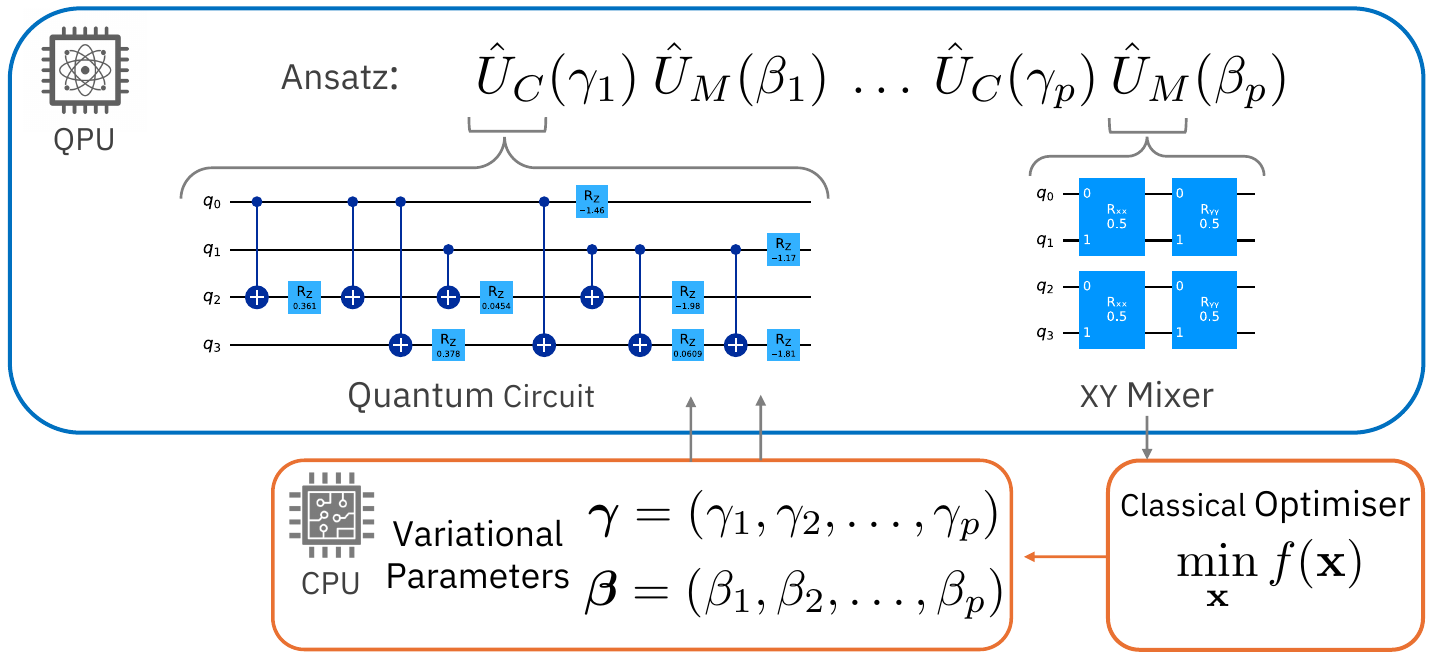}
     \caption{Schematic of the hybrid classical-quantum workflow of \gls{QAOA} with $p$ layers. The quantum circuit composing the cost Hamiltonian and the custom XY mixer are represented, as well as the classical optimiser, \gls{COBYLA}, that updates the variational parameters at each iteration.}
     \label{fig:QAOA_workflow}
\end{figure}

The Quantum Approximate Optimisation Algorithm~\cite{farhi2014quantum} is a hybrid quantum-classical algorithm designed to find approximate solutions to \gls{CO} problems. The \gls{QAOA} is an example of a variational quantum algorithm with a fixed form for the ansatz. As shown in Fig. \ref{fig:QAOA_workflow}, the \gls{QAOA} involves alternating applications of a parameterised cost unitary $\hat{U}_C(\gamma)$ and mixer unitary $\hat{U}_M(\beta)$. The cost and mixer unitaries are repeated $p$ times, giving a circuit of the form,
\begin{equation}
	\hat{U} = \Pi_{i=1}^p \hat{U}_M(\beta_i) \hat{U}_C(\gamma_i),
\end{equation}
which has a total of $2p$ variational parameters that are optimised by a classical optimiser. The cost unitary is $\hat{U}_C(\gamma) = e^{-i\gamma\hat{H}_C}$ where $\hat{H}_C$ is the cost Hamiltonian, i.e. the qubit form of Equation \eqref{eq:Ising-Ham}. The mixer unitary has the form $e^{-i\beta\hat{H}_M}$ where $\hat{H}_M$ is a mixer Hamiltonian designed to enable exploration of the solution space~\cite{blekos2023review}. Conceptually, the form of the ansatz is heavily inspired by annealing, with a combination of time evolution and `hopping'. Given that $\hat{H}_C$ encodes the cost function of the optimisation problem, time evolution under this $\hat{H}_C$ alone would lead to trivial dynamics because energy (cost) would be conserved and therefore no choice of $\gamma$ would improve the cost. The mixer Hamiltonian is chosen so that it does not commute with $\hat{H}_C$ and therefore leads to non-trivial dynamics (effectively enabling a `hopping' between states) as the cost is no longer conserved.

There are several possible choices of mixer Hamiltonians, the simplest of which is the transverse field mixer, $\hat{H}_M = \sum_i X_i$. Another choice is the so-called XY-mixer~\cite{Fuchs_2022} which has the property that, in the noiseless case, it conserves the Hamming weight of the input state. We refer to the \gls{QAOA} combined with the transverse field mixer or the XY-mixer as \gls{QAOA} or XY-QAOA, respectively. This is particularly useful in the present because the physical states all have a fixed Hamming weight. The XY-mixer therefore ensures that the hopping sends constraint-satisfying states to other constraint-satisfying states, and is therefore a hard constraint method. This is in contrast to the transverse field mixer, which does not preserve Hamming weight and would therefore hop from valid bitstrings to invalid bitstrings. The standard form of the XY-mixer is,
\begin{equation}
	\hat{H}_M^{XY} = \frac{1}{2}\sum_{i=0}^{M-1} (X_i X_{i+1} + Y_i Y_{i+1}),
\label{eq:XY-global}
\end{equation}
where $X_M=X_0$ and $Y_M=Y_0$. The standard XY-mixer preserves the Hamming weight globally, i.e. the Hamming weight of the full bitstring is preserved. Note that when the index goes above $M$, it wraps back to 1. However, the structure of our problem means that while the conventional XY-mixer would preserve the global Hamming weight of $N$, it would allow transitions to invalid bitstrings that do not preserve local Hamming weight constraints. As such, we use an alternate, local form of the XY-mixer defined as,
\begin{equation}
\hat{H}_M^{XY} =
\begin{cases}
\begin{aligned}[t]
    &\frac{1}{2} \sum_{i=0}^{N - 1} \sum_{j=0}^{n-1} \big(
    X_{in + j} X_{in + (j+1) \bmod n} \\
    &+\,Y_{in + j} Y_{in + (j+1) \bmod n},
    \quad \hspace{10pt} \scriptstyle n > 2,\, N > 2
\end{aligned} 
\\[8pt]
\\[1pt]
\frac{1}{2} \left(X_0 X_1 + Y_0 Y_1 \right), \quad \hspace{37pt} \scriptstyle n=2,\,N=2
\end{cases}
\label{eq:XY-local}
\end{equation}
where the inner index $j$ runs over qubits within each residue block of size $n$, and the term $(j+1) \bmod n$ ensures that the last qubit in each block interacts with the first, to give a \textit{ring}-XY model that includes one-dimensional nearest-neighbour interactions with a periodic boundary condition \cite{heAlignmentInitialState2023}. The local version of the XY-mixer has the benefit that, in the noiseless case, it only allows transitions between valid states. The local version of the XY-mixer has the benefit that, in the noiseless case, it only allows transitions between valid states.

While the XY-mixer is a useful tool, it comes at the cost of increased circuit depth (see section \ref{sec:circuit_depth}). As such, in the near term, the extra noise introduced by the increased depth may offset any advantage from the XY-mixer. This may especially be the case because in the presence of noise the XY-mixer is no longer guaranteed to preserve Hamming weight. There have been studies comparing both the effectiveness and susceptibility to noise of the transverse field and XY-mixers \cite{Niroula_2022}.

\section{Computational Details}
\label{sec:computational_details}
Given a protein with  $N$ residues and  $n$ rotamers per residue, the accessible Hilbert space of this system contains $n^N$ bitstrings. As the evaluation of the energy of each bitstring corresponds to a matrix multiplication of dimension $M=Nn$, the exact ground state energy of this system can be found in $\mathcal{O}(N^2 n^{N+2})$ by a brute-force search over the allowed bitstrings. We evaluate the exact ground state energies this way and use these energies as a reference, noting that this is only possible for smaller system sizes.

In this work, we make use of two simulators from Qiskit Aer to perform XY-QAOA simulations. For systems up to 28 qubits, we simulate the algorithm exactly using the sampler primitive based on the statevector simulator, which we refer to as \gls{SV-QAOA}. To extend beyond these relatively small-scale simulations, we also perform experiments based on Qiskit's \gls{MPS} simulator, hereafter referred to as \gls{MPS-QAOA} for systems with 5 and 6 residues, up to 54 qubits. Unlike \gls{SV} simulations, which track the full $2^M$-dimensional quantum state, the \gls{MPS} representation approximates the state as a product of low-rank tensors, drastically reducing memory requirements. This approximation is accurate as long as the entanglement across the system remains low. 

\begin{figure}[b]
\centering
\resizebox{\linewidth}{!}{%
\begin{tikzpicture}
\node (left) at (0.5,0) {
\begin{quantikz}
& \gate[2]{A(\theta,\phi)} & \qw  \\
& & \qw
\end{quantikz}
};

\node (eq) at (2.2,0) {\Large$=$};

\node (right) at (6.5,0) {
\begin{quantikz}
& \targ{} & \qw & \ctrl{1} & \qw & \targ{} & \qw \\
& \ctrl{-1} & \gate{R(\theta,\phi)^\dagger} & \targ{} & \gate{R(\theta,\phi)} & \ctrl{-1} & \qw
\end{quantikz}
};
\end{tikzpicture}
}
\caption{Decomposition of the \( A(\theta, \phi) \)-gate into elementary gates. \( R(\theta, \phi) = R_z(\phi+\pi) R_y(\theta+\pi/2) \) where \( R_z(\theta) = \exp{(-i\theta\sigma_z/2)}\) and \( R_y(\theta) = \exp{(-i\theta\sigma_y/2)}\). $\sigma_z$ and $\sigma_y$ are Pauli matrices. The simulations were performed with $\theta=\pi/4$ and $\phi=0$.}
\label{fig:Agate}
\end{figure}
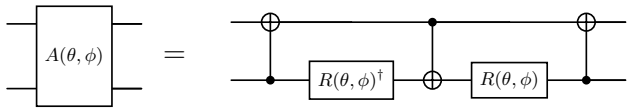

The performance of \gls{QAOA} depends on the choice of the initial state. Z. He et al. \cite{heAlignmentInitialState2023} showed that initialising \gls{QAOA} with a state of high fidelity to the ground state of the XY-mixer leads to higher approximation ratios. Warm-starting strategies, where a classical algorithm produces a high-performing \gls{QAOA} initial state, have also been proposed \cite{Egger_2021, tate2023bridging}. By ensuring that the initial state is a superposition of Hamming weight preserving bitstrings, we can limit the \gls{QAOA} to sample only the allowed bitstring subspace \cite{fingerhuth2018quantum}. The ideal case is to begin with an initial state that is a uniform superposition over bitstrings of required Hamming weight—a so-called Dicke state~\cite{PhysRev.93.99}. However, Dicke states can be non-trivial to prepare, and must be implemented via a circuit of $\mathcal{O}(nN)$ depth~\cite{2019andreas}. We prepare the \gls{QAOA} initial states as superpositions of states (not necessarily uniformly) with the required Hamming weight $N$. First, we construct a subcircuit of $n$ qubits to represent a single residue with $n$ rotamers. We apply an $X$ gate to one of the qubits so as to promote this subsystem to the subspace with the correct Hamming weight. Next, a cascade of custom $A$ gates (as defined by Equation~(2) of ref~\cite{Gard_2020} with $\theta=\pi/4$ and $\phi=0$, see Fig. \ref{fig:Agate}) is applied to the nearest-neighbour qubits, which mixes the Hamming weight preserving states. Finally, the full quantum circuit corresponding to the initial state of the system is constructed as a product state of the circuits corresponding to the individual substrings. An initial state prepared this way has a depth of $\mathcal{O}(n)$.

Finding an optimum set of parameters for variational algorithms like \gls{QAOA} is an NP-hard problem~\cite{Akshay2020, Bittel2021}. The convergence of \gls{VQAs} are known to depend heavily on the choice of initial parameters~\cite{Anschuetz_2022}, and methods have been proposed to determine better initial points~\cite{Yang_2024}. We perform several trajectories of each simulation with randomised initial points and average over the results. We initialise the cost unitaries by drawing parameters $\gamma_i$ uniformly from the range $[-0.1,0.1]$ and the mixer unitaries by drawing parameters $\beta_i$ uniformly from $[-1.0,1.0]$ (angles are in radians). These values are then merged by alternating elements to give the initial point. This choice promotes the mixing of Hamming weight preserving states and aims to avoid encountering local minima early in the optimisation.

Each XY-QAOA simulation consists of iteratively sampling the ansatz for a fixed number of shots and optimising the ansatz parameters, until convergence is reached. For smaller systems, the number of shots for \gls{SV-QAOA} is varied in the order of 10-100 to obtain meaningful statistics, and the \gls{QAOA} depth is fixed at $p=4$. Using a larger number of shots fixed across all \gls{SV-QAOA} simulations would lead to a constant scaling of the number of total shots required to find the ground state with respect to system size, since at the system sizes accessible to \gls{SV-QAOA}, the ground state would almost always be found within the first iteration of the algorithm. \gls{SV-QAOA} simulations are averaged over up to 20 trajectories, all of which converge to the ground state. In the case of \gls{MPS-QAOA}, we select $p = 4$ for 5-residue systems, whereas for 6-residue systems, we set $p = 25$. This increase in $p$ is motivated by the observed improvement in convergence ratios, defined as the fraction of trajectories that successfully reach the ground state of the system. This is consistent with the expectation that circuits with higher expressivity are needed to capture the increased complexity of larger systems. However, we note that increasing $p$ could lead to barren plateaus~\cite{Larocca2022diagnosingbarren,zhang2022quantum}. The number of shots per iteration is set to 1000 for all \gls{MPS-QAOA} simulations. While our choices of hyperparameters yield satisfactory convergence (see Appendix \ref{app:mps_sa_success}), further improvements may be reached by fine-tuning these parameters.

We employ the classical gradient-free optimiser \gls{COBYLA} solver~\cite{2020SciPy-NMeth, sturm2023theory}, to optimise the variational parameters. We also use the Conditional Value at Risk (CVaR) method~\cite{Barkoutsos2020improving}, which considers only the $\alpha$ tail of the distribution rather than the full distribution when optimising the variational parameters. In our numerical experiments, we set a value of $\alpha = 0.2$, i.e. the optimisation is driven by the 20\% tail of the distribution. 

For the classical experiments, we employ an improved version of \gls{SA} known as the dual annealing algorithm. Dual annealing is a hybrid global optimisation method that combines \gls{SA} for global exploration with a local search phase to refine candidate solutions, typically using a gradient-based method such as L-BFGS-B \cite{lbfgsb, 10.1145/279232.279236}. We use the implementation in the \texttt{optimize} module from the \texttt{Scipy} library~\cite{2020SciPy-NMeth,RJ-2013-002}, which in turn applies the Powell algorithm. The hyperparameter configuration $\texttt{visit} = 1.01$ and $\texttt{accept} = 0.9$, was found to outperform the default values of $\texttt{visit} = 2.62$ and $\texttt{accept} = -5$, which are based on Tsallis statistics \cite{Tsallis}. With 1000 optimisation iterations, this refined setting yielded more consistent convergence to the ground state and reduced the total number of function evaluations required.

\section{Results and Discussions}
\label{sec:results}
\subsection{Implementation of the Constraints}
\label{sec:circuit_depth}
The choice of method for implementing constraints is influenced by several factors. In principle, one would opt for the choice that maximises the in-constraint ratio, i.e. the ratio of valid to constraint-violating solutions. In the absence of noise, this will be a hard-constraint method such as XY-QAOA. However, in practice all methods for enforcing constraints other than post-selection will increase circuit depth. In the near-term, the increased noise due to larger circuit depth may destroy any gains that would have been seen in the noiseless case. This behaviour was observed in the work by Niroula et al. \cite{Niroula_2022}. Post-selection, by contrast, refers to employing the cost Hamiltonian without penalty terms and the transverse field mixer, and selecting only the bitstrings that satisfy the constraints. 

Each of the choices involves trade-offs. For example, post-selection does not introduce overhead in terms of circuit depth, which is critical in the near-term. However, it will introduce a large (possibly combinatorial) sampling overhead. Both soft and hard constraints in the form of local penalties and the XY-QAOA will introduce some circuit depth overhead but will increase the probability of sampling constraint-satisfying solutions. Penalty terms can complicate the optimisation process, potentially leading to trade-offs that form a Pareto frontier~\cite{wang2025mmdnewtonmethodmultiobjectiveoptimization}.

\begin{table}[t]
    \centering
    \small
    \setlength\tabcolsep{1pt}
    \renewcommand{\arraystretch}{1.}
    \resizebox{\linewidth}{!}{%
    \begin{tabular}{|c|c|c|c|c|c|c|c|}
        \hline
        {Rot.} & {Res.}  & {Method} & {CD} & {CD-SP} & {T-CD} & {Opt T-CD} & {Opt CNOTs}\\
        \hline \hline
        \multirow{3}{*}{2} & \multirow{3}{*}{2} 
        & XY-QAOA & 6 & 9 & 25 & 18 & 27 \\
        & & pen-\gls{QAOA} & 6 & 6 & 20 & 14 & 17 \\
        & & Baseline & 4 & 4 & 15 & 11 & 14 \\
        \hline
        \multirow{3}{*}{3} & \multirow{3}{*}{3} 
        & XY-QAOA & 22 & 28 & 76 & 64 & 156 \\
        & & pen-\gls{QAOA} & 20 & 20 & 75 & 47 & 107 \\
        & & Baseline & 16 & 16 & 52 & 35 & 93 \\
        \hline
        \multirow{3}{*}{4} & \multirow{3}{*}{4} 
        & XY-QAOA & 20 & 29 & 202 & 168 & 544 \\
        & & pen-\gls{QAOA} & 22 & 22 & 220 & 158 & 434 \\
        & & Baseline & 16 & 16 & 147 & 124 & 412 \\
        \hline
        \multirow{3}{*}{5} & \multirow{3}{*}{5} 
        & XY-QAOA & 34 & 46 & 313 & 266 & 991 \\
        & & pen-\gls{QAOA} & 30 & 30 & 400 & 305 & 939 \\
        & & Baseline & 28 & 28 & 231 & 166 & 734 \\
        \hline
        \multirow{3}{*}{6} & \multirow{3}{*}{6} 
        & XY-QAOA & 36 & 51 & 521 & 428 & 1809 \\
        & & pen-\gls{QAOA} & 42 & 42 & 652 & 434 & 1528 \\
        & & Baseline & 32 & 32 & 363 & 340 & 1352 \\
        \hline
        \multirow{3}{*}{7} & \multirow{3}{*}{7} 
        & XY-QAOA & 40 & 58 & 737 & 531 & 2834 \\
        & & pen-\gls{QAOA} & 46 & 46 & 863 & 692 & 2633 \\
        & & Baseline & 34 & 34 & 662 & 548 & 2454 \\
        \hline
    \end{tabular}}
    \caption{Table of circuit depths—defined as the number of layers of CNOT gates—for different numbers of rotamers per residue with three approaches: XY-QAOA (\gls{QAOA} with constraints enforced via the XY mixer), pen-QAOA (\gls{QAOA} with constraints enforced via local penalty terms and a standard X mixer), and a baseline method (\gls{QAOA} without constraint enforcement, using a standard X mixer). For each method, we report: the logical circuit depth (CD) and the logical circuit depth including state preparation (CD-SP); the post-transpilation metrics, including state preparation, for circuits transpiled to \texttt{ibm\_torino}: the transpiled circuit depth without optimisation (T-CD); and the optimised transpiled circuit depth (Opt T-CD) and CNOT count (Opt CNOTs) after applying Qiskit transpiler optimisation. The results shown are for QAOA with one ansatz layer.}
    \label{table:circuit_depths}
\end{table}

In Table \ref{table:circuit_depths} we compare the pre-compiled logical circuit depth (CD) defined as the number of layers of CNOT gates for each of the constraint methods, for a QAOA ansatz with one layer. The logical circuit depth corresponds to the minimum number of layers required, assuming optimal grouping of all mutually commuting two-qubit Pauli terms into parallelisable layers. Specifically, we compare the depth of the local penalty method using a standard transverse field mixer Hamiltonian (pen-QAOA) with that of the XY-mixer method (XY-QAOA). We also evaluate a baseline case (post-selection), which has no penalty terms and the transverse field mixer. For the baseline method, the logical circuit depth is twice the number of non-commuting $ZZ$ gate layers arising from the cost Hamiltonian defined in Equation~\eqref{eq:Q_matrix}. Local penalties are incorporated as additional non-zero $ZZ$ interactions between rotamers on the same residue in the cost Hamiltonian. 

The local XY-mixer as defined in Equation~\eqref{eq:XY-local} introduces one $XX+YY$ interaction between each pair of neighbouring qubits within a block of size $n$, including a term connecting the last and first qubit of each residue block to enforce periodicity. Each $XX+YY$ term, also known as an $XY$ gate, can be implemented using two CNOT gates.  Due to the disjoint block structure, terms in different blocks commute and can be scheduled independently. Within each block, the scheduling problem reduces to edge colouring a ring of size $n$, which requires either two or three layers depending on whether $n$ is even or odd, respectively. Thus, the CNOT depth increases due to the XY mixer alternates between 4 and 6 for even and odd $n$ respectively, except in the minimal case $(n=2, N=2)$ which only adds 2 CNOT layers with respect to the baseline method. 

We note that for a given number of rotamers per residue $n$, CD tends to plateau as the number of residues $N$ increases. This is because the two-qubit unitaries are highly parallelisable over residues, thereby preventing a proportional increase in circuit depth. As the Table~\ref{table:circuit_depths} shows the CD for a single \gls{QAOA} layer, for a depth $p$ ansatz, the corresponding CD will be $p$ times the value for a single \gls{QAOA} layer. 

We additionally report the pre-compiled logical circuit depth when accounting for initial state preparation (CD-SP), which adds $3(n-1)$ entangling gates to the XY-QAOA circuit, where $n$ is the number of rotamers per residue. In contrast, the baseline and local penalty QAOA initialise the circuit using a simple bitstring-dependent layer of single-qubit $X$ gates, which does not contribute to the entangling gate depth. We also report the entangling gate depths of the circuits transpiled to the 133-qubit \texttt{ibm\_torino} device with Heron r1 architecture. Here, T-CD indicates the unoptimised circuit depth, while Opt T-CD and Opt CNOTs represent the circuit depth and CNOT count, respectively, after optimising with Qiskit's transpilers.

As expected, Table \ref{table:circuit_depths} shows that the post-selection method achieves the lowest depth, both for the logical and transpiled circuits (CD and T-CD). The hard-constraint method yields lower CD compared to the soft-constraint approach, especially as system size increases. This is because the XY mixer applies only local, nearest-neighbour two-qubit interactions within residue blocks, which can be efficiently parallelised. In contrast, the soft-constraint approach adds $ZZ$ interaction terms between all qubits within a block, increasing the number of non-commuting terms and thus the circuit depth required for their implementation. However, once the depths of the state preparation circuits are taken into account, CD-SP for the XY-QAOA method exceeds that of the penalty method. Nevertheless, in the limit of large $p$, the state preparation cost will become insignificant relative to the cost of the \gls{QAOA} ansatz. The transpiled circuit depth (T-CD) shows fewer CNOT gate layers for the XY-QAOA method for large system sizes, where the circuit structure has stabilised and the larger variability at small system sizes is surpassed. With Qiskit transpiler optimisation techniques applied, the hard constraint method exhibits an even greater reduction in circuit depth (Opt T-CD). We anticipate that additional optimisation techniques could further reduce circuit depths beyond the levels achieved here, as our current results are based solely on Qiskit's transpiler methods.

\subsection{Scaling of the Classical and Quantum Methods}
\label{sec:scaling}
Our goal is to compare the scaling of our \gls{QAOA} algorithm proposed, to a classical algorithm in the setting of finding the exact ground state of the protein side-chain optimisation problem. We define computational cost as the minimum number of function calls required by either the classical \gls{CPU}-based solver or the \gls{QPU}-based method to reach the exact ground state. While classical solvers such as CPLEX \cite{cplex2009v12} or Gurobi \cite{gurobi} are among the most well-established and reliable tools for combinatorial optimisation, their scalability can become a limitation for larger problem instances. For this reason, we choose to compare against \gls{SA} as it is a competitive classical method, and as a heuristic it is conceptually similar to the \gls{QAOA}. Although the hyperparameters of the \gls{SA} algorithm were tuned to enhance performance in our experiments, we do not assert that these values are universally optimal. The outcomes reported are inherently dependent on this choice, and different settings may lead to variations in convergence behaviour and solution quality. Given that we are looking for an exact solution to an NP-hard problem, it is expected that both classical and quantum methods will scale exponentially with problem size. Nonetheless, it is still instructive to attempt to estimate if there is a crossing point where the quantum method begins to outperform the classical method and if so, at what scale this occurs. Such numerical experiments can help to inform expectations about the future utility of the quantum approximate optimisation algorithm.

\begin{figure}[b]
    \centering
    \includegraphics[width=\linewidth]{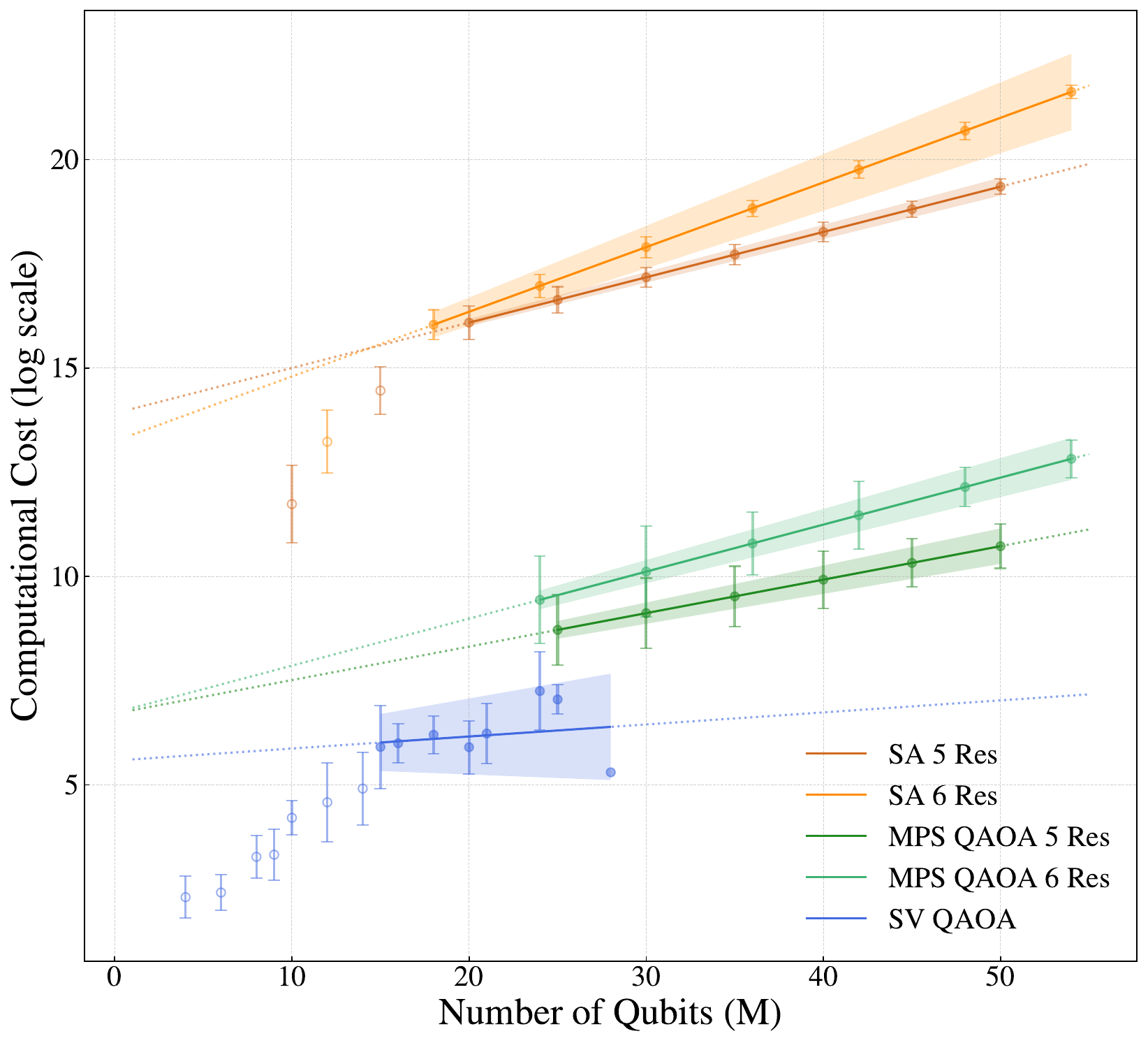}
    \caption{The computational cost—defined as the number of calls to the \gls{CPU} or \gls{QPU}, respectively, required to find the ground state—for the \gls{SA} and \gls{QAOA} optimisation algorithms. Results are presented on a semi-log plot, showing the mean and standard deviation for both trends. \gls{SV-QAOA} simulations are shown in blue, and \gls{MPS-QAOA} simulations in green.}
    \label{fig:cpu_vs_qpu}
\end{figure}

To compare the computational cost of the quantum and classical algorithms, we define the computational cost as follows. In the classical setting, the leading term in the computational cost is proportional to the number of times the energy evaluation function is called in the subroutine. Whereas, the quantum computational cost is found by simply counting the total number of quantum circuits executed to find the ground state. For both classical and quantum cases, the computational cost is normalised by dividing it by the corresponding convergence ratio. Note that we stop the quantum algorithm the first time it sees the ground state even if this occurs during the training process. In other words, we do not necessarily wait until the variational parameters converge to stop the algorithm. This is justified because the \gls{QAOA} is a sampling algorithm rather than an expectation value algorithm, and so in practice all that is needed is to see the ground state once. \\

In Fig.~\ref{fig:cpu_vs_qpu}, we show the scaling trends in computational cost for both the classical and quantum methods used to find the ground state. 
We first plot the data on a logarithmic scale and then perform a linear regression to extract the scaling behaviour of the \gls{SA} and \gls{QAOA} algorithms. The fits are performed starting from $M=18$ qubits for both \gls{SA} and \gls{MPS-QAOA}, and from $M=15$ qubits for \gls{SV-QAOA}, as smaller system sizes follow a pre-asymptotic regime. 

The results demonstrate distinct exponential scaling behaviours in the overall computational cost. For the classical \gls{SA} method (orange), the cost scales approximately as $\mathcal{O}(e^{A M})$, with parameters $A=0.109 \pm 0.004$ for systems with 5 residues, and $A=0.155 \pm 0.017$ for 6 residues. In contrast, the quantum approaches exhibit significantly milder exponential scaling. The \gls{SV-QAOA} method (blue) shows scaling with $A=0.029 \pm 0.046$, while the \gls{MPS-QAOA} approach yields  $A=0.080 \pm 0.009$ for 5 residues, and $A=0.113 \pm 0.009$ for 6 residues. The corresponding $r^2$ values for all fits are provided in Appendix~\ref{app:mps_sa_success}.

There is clearly a vertical offset between the \gls{SV-QAOA} and \gls{MPS-QAOA} methods, reflecting a larger constant prefactor in the scaling relation. This derives from the choice in number of shots per iteration, which for \gls{MPS-QAOA} was set to $1000$, whereas for the \gls{SV-QAOA} simulations was in the order of $10 \text{-} 100$. However, what is relevant here is the scaling behaviour of the methods. As previously discussed, to account for the fraction of converging trajectories in the algorithms analysed, we divide the computational cost by the corresponding convergence ratio. This correction has no effect on the \gls{SV-QAOA} results, which always converge, whereas both \gls{MPS-QAOA} and \gls{SA} are affected due to decreasing convergence ratios with increasing system size. Before applying the convergence correction, the two quantum methods exhibit approximately parallel slopes, indicating similar scaling with system size—a promising sign of their mutual consistency, even if the data for \gls{SV-QAOA} appears somewhat more variable. After normalising by the convergence ratio, the slope for \gls{MPS-QAOA} increases, reflecting the non-convergence of the approximate \gls{MPS} method.
In our subsequent analyses, we will consider the \gls{MPS-QAOA} fit as the primary indicator of quantum scaling behaviour, while treating the \gls{SV-QAOA} results as a proof of concept. 

We note here that the effect of training the parameters of the \gls{QAOA} algorithm also accounts for a portion of the shots necessary to reach the ground state and so the computational cost. By warm-starting the \gls{QAOA} simulations with the optimal parameters from the trained circuit, the computational cost of the quantum optimisation can be reduced further~\cite{Egger_2021}.

\begin{figure}[t]
    \centering
    \includegraphics[width=\linewidth]{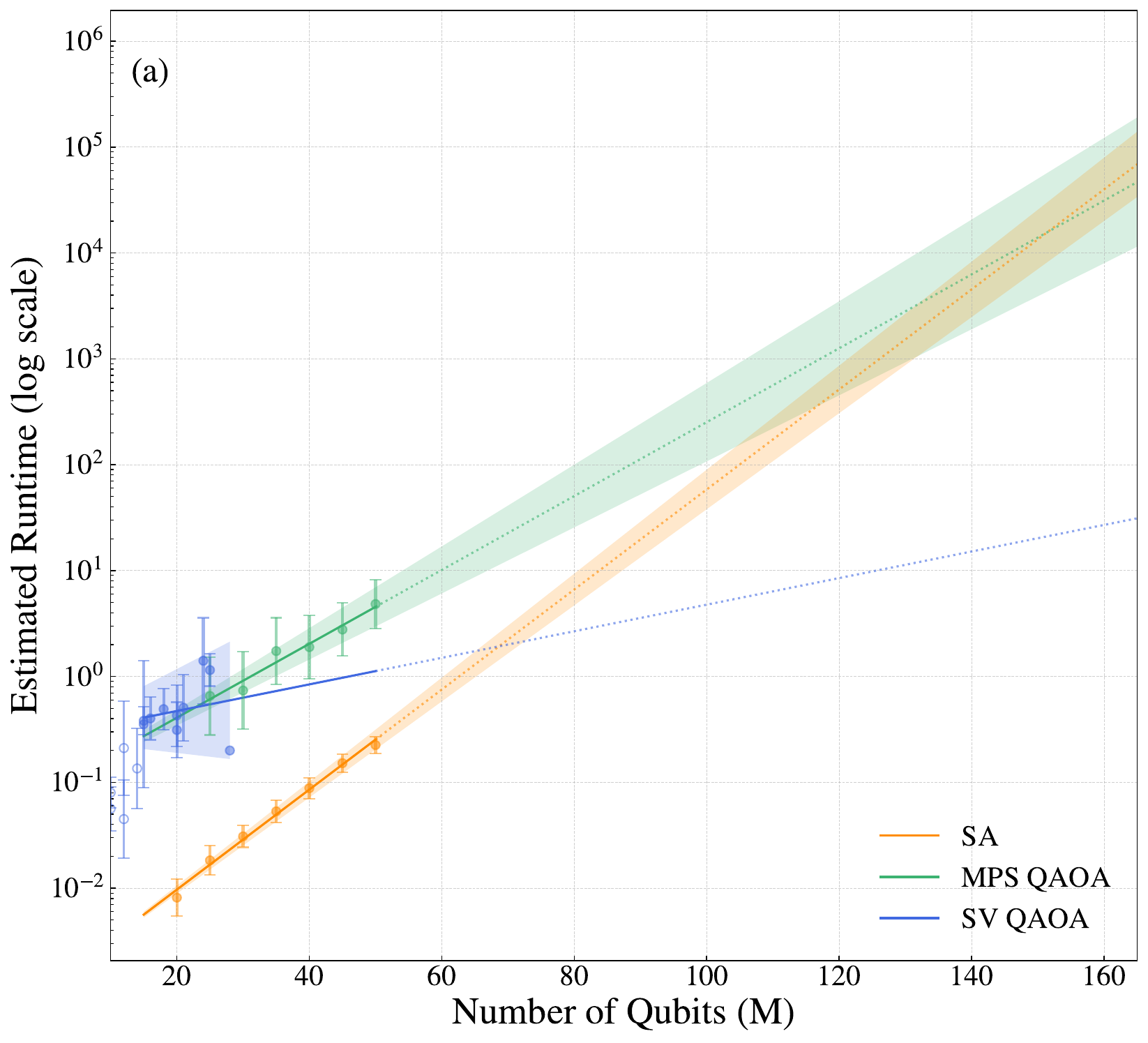}
    
    \vspace{0.5em}
    
    \includegraphics[width=\linewidth]{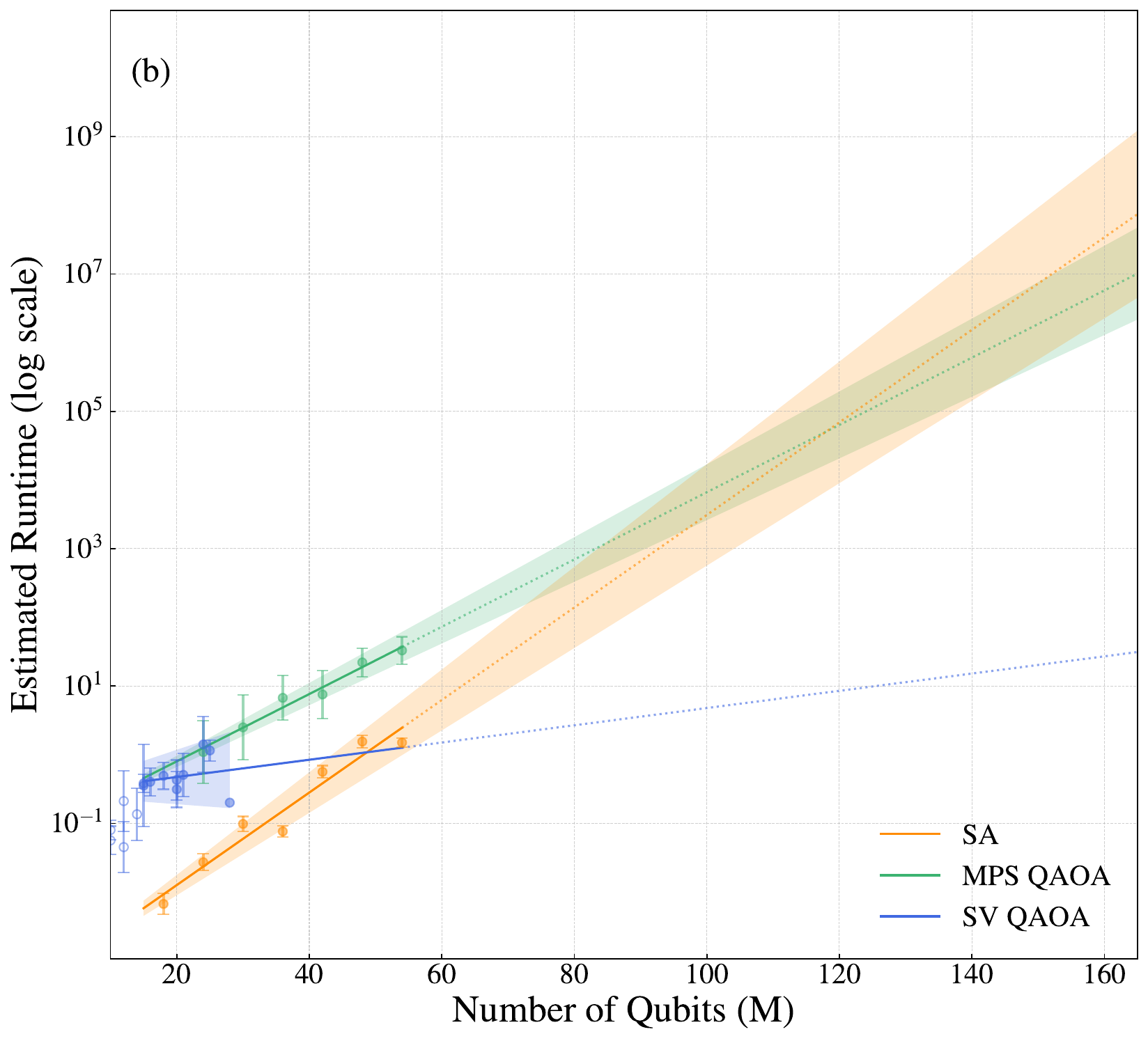}
    \caption{Estimated crossover point between the runtime of the classical \gls{SA} algorithm—normalised by the order of magnitude of the average \gls{CPU} clock speed (1~GHz)—and the quantum \gls{QAOA} optimisation algorithm—normalised by the order of magnitude of the average \gls{QPU} clock speed (1~kHz) for 5 residues in (a) and 6 residues in (b). The shaded regions represent the uncertainty in the gradients of the \gls{SA} and \gls{MPS-QAOA} slopes, with the latter selected rather than \gls{SV-QAOA}, for its greater stability in gradient estimation.}
    \label{fig:runtimes}
\end{figure}

As mentioned previously, we interrupt both the classical and the quantum algorithms upon the first occurrence of the exact ground state. Naturally, this is only feasible while the system size remains small enough so that the true ground state can be found. Once the system scales beyond this point, early termination is no longer possible for any of the algorithms—classical or quantum. It remains speculative whether the trends identified in systems below 50 qubits will persist as the system size increases. Nonetheless, simulated annealing (SA) samples from the full $2^{Nn}$ configuration space, which includes many invalid states. Without early termination of the optimisation, it is likely to spend considerable time exploring these invalid or suboptimal regions. In contrast, our \gls{QAOA} algorithm is designed to concentrate sampling within the valid $n^N$ subspace and so even without early termination, sampling will be more concentrated on valid configurations. Therefore, in the worst case, we expect the scaling behaviour of the algorithms to remain consistent.

It is important to note that our study emphasises the scaling behaviour of quantum and classical algorithms, rather than their absolute wall-time performance which is influenced by constant pre-factors. Given that \gls{QPU}s typically operate at frequencies in the kilohertz ($kHz$) range, in contrast to the gigahertz ($GHz$) frequencies common in \gls{CPU}s, we anticipate that these pre-factors will shift the cost curve upward for quantum algorithms relative to their classical counterparts. As demonstrated, the classical \gls{CPU}-based approach requires nearly five orders of magnitude more function evaluations than its quantum counterpart for larger problem sizes, a disparity that continues to widen with increasing system size. This pronounced difference in computational cost underscores the superior asymptotic scaling potential of the quantum algorithm. Improved scaling suggests that, as problem sizes grow, quantum methods may not only surpass classical approaches but also remain competitive where classical methods become increasingly inefficient. Moreover, ongoing advancements in quantum hardware could accelerate the realisation of these theoretical speed-ups in practical applications. 

In Fig.~\ref{fig:runtimes}, we estimate the crossover point between the runtime costs of the classical and quantum optimisation algorithms. This is achieved by normalising the computational cost by the average clock speeds of the \gls{CPU} ($GHz$) and the \gls{QPU} ($kHz$), providing an approximate comparison of their respective runtimes. Based on this estimation, and considering current quantum hardware capabilities, the crossover point is projected to occur at a problem size of approximately 115 to 160 qubits.

The shaded regions in the plot represent the variability in slope estimation (on the logarithmic scale) for both the \gls{QPU} runtime, approximated using the \gls{MPS} simulator, and the \gls{SA} runtime. These error bounds reflect uncertainty in the crossover point, which may shift depending on the evolving performance of the respective computational platforms. In a conservative scenario, assuming worst-case performance for \gls{MPS} and best-case for \gls{SA}, the crossover could occur before reaching 315 qubits. It is important to emphasise that this estimate is based on extrapolated fits for the particular problem modelling approach chosen here, and should therefore be interpreted as indicative rather than definitive. Additionally, variations in the effective \gls{QPU} clock speed, which are not accounted for in this plot, could lead to an earlier crossover in the case of faster quantum devices. As quantum hardware continues to advance, further deviations from the projected trends are also plausible. 

\section{Conclusion}
In this work, we present a Qiskit-based pipeline for solving the protein side-chain optimisation problem, focusing on the internal degrees of freedom—rotamers—given a fixed protein backbone. By deriving a Quadratic Unconstrained Binary Optimisation (QUBO) model from the physical formulation and mapping it to an Ising model, we enable quantum encoding of the problem. This sets the stage for quantum optimisation by means of our proposed Quantum Approximate Optimisation Algorithm (QAOA). To enforce the necessary constraints, we explore both soft penalty-based methods and a hard-constraint approach using the XY mixer. Our results highlight the resource efficiency of the XY-QAOA variant, particularly in terms of reduced circuit depth and fewer CNOT gates, while imposing the Hamming's condition by definition. We benchmark the performance of our XY-QAOA algorithm against a classical heuristic—Simulated Annealing (SA)—by comparing their ability to find the exact ground state. We evaluate this in terms of computational cost, defined as the number of calls to the quantum processing unit (QPU) or, respectively, the classical processing unit (CPU), required to reach the ground state for the first time. A flattening trend in the quantum method’s scaling, observed on a semi-logarithmic plot, suggests favourable asymptotic behaviour as system size increases. To provide a practical perspective, we estimate a crossover point between the quantum and classical methods by normalising for the current average clock speeds of the \gls{CPU} and \gls{QPU}. Our analysis suggests an indicative crossover point at around 115 to 150 qubits—given current quantum hardware capabilities—where the quantum optimisation algorithm could potentially outperform its classical counterpart. While we do not claim that \gls{SA} is the highest-performing classical heuristic for this problem, ongoing advancements in quantum hardware suggest that clock speeds are likely to improve beyond their current values, meaning that the crossover point identified here may shift further. Nevertheless, we should note that noise and error correction techniques have not been accounted for here, which would detrimentally affect the quantum computations.

In this study, the selection of the rotamers modelled to the structure was performed arbitrarily. A more systematic approach to rotamer selection should be considered—for instance, based on energy minimisation or a geometry-based criterion, such as selecting rotamers that maximise the physical space covered. Furthermore, incorporating three-body interaction terms into the model would provide a more accurate representation of the energy landscape. Finally, applying this methodology to real protein structures of physical relevance—such as the MV1-linker peptide (an apoptosis inhibitor) or Macimorelin (a drug for the diagnosis of adult growth hormone deficiency)—would offer additional validation and demonstrate the applicability of the proposed approach in real-world scenarios.

\section*{Data and code availability}
We provide a publicly accessible GitHub repository (\href{https://github.com/stfc/quantum-protein-folding}{github.com/stfc/quantum-protein-folding}) containing the data, results, and code associated with this manuscript.

\vspace{5pt}

\section*{Acknowledgements}
The authors thank Benjamin C. B. Symons for fruitful discussions. We are grateful for the support from the NCCR MARVEL, a National Centre of
Competence in Research, funded by the Swiss National Science Foundation (grant number 205602). This work was supported by the Hartree National Centre for Digital Innovation, a UK Government-funded collaboration between STFC and IBM. IBM, the IBM logo, and ibm.com are trademarks of International Business Machines Corp., registered in many jurisdictions worldwide. Other product and service names might be trademarks of IBM or other companies. The current list of IBM trademarks is available at \href{https://www.ibm.com/legal/copytrade}{www.ibm.com/legal/copytrade}

$\,$

$\,$
\bibliography{references}
$\,$
$\,$

\newpage
\appendix
\refstepcounter{section}
\section*{Appendix \Alph{section}: Rotamer Interactions}
\addcontentsline{toc}{section}{Appendix \Alph{section}: Rotamer Interactions}
\label{app:rot_int}

\begin{figure}[b]
    \centering
    \includegraphics[width=\linewidth]{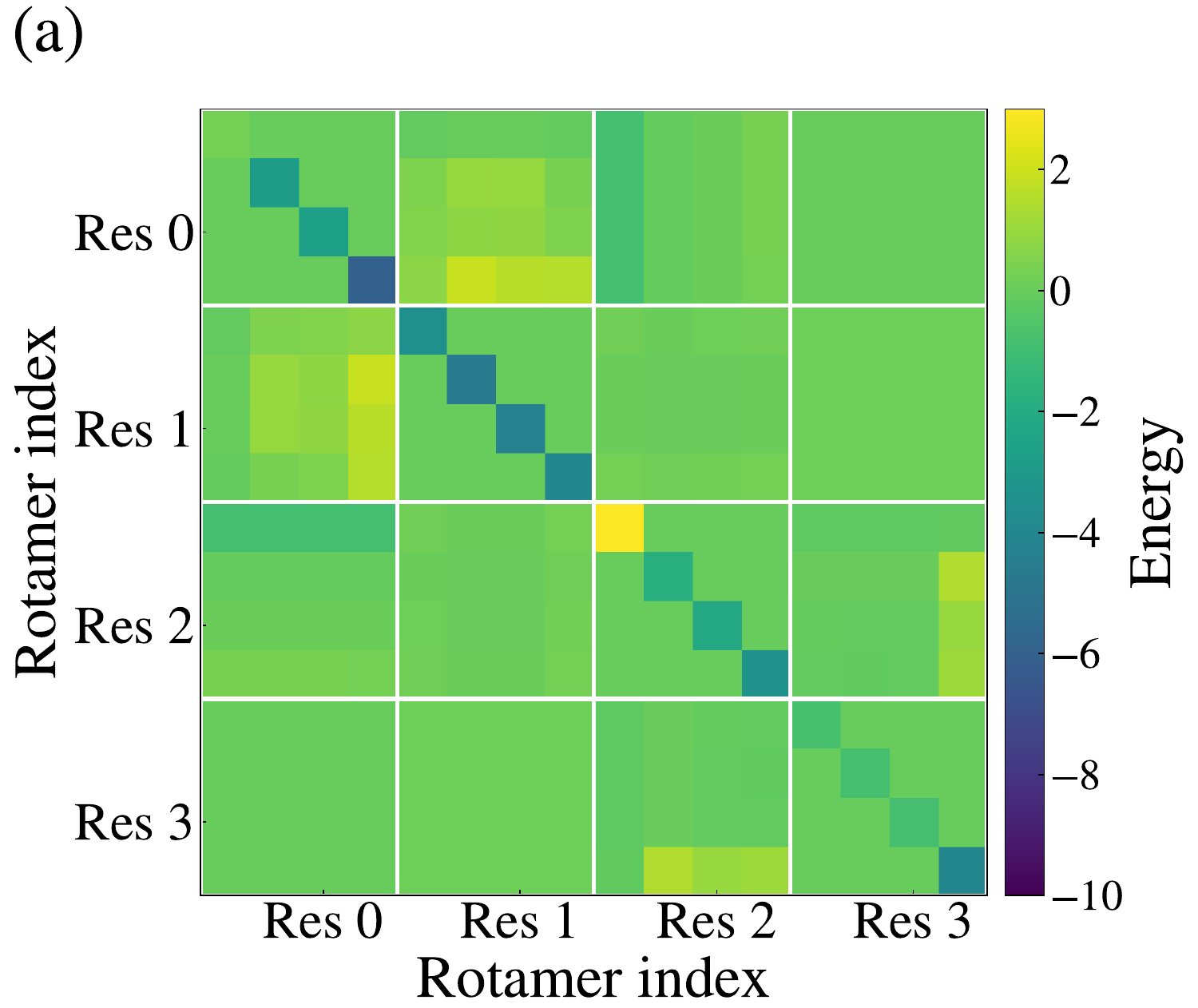}
    
    \includegraphics[width=\linewidth]{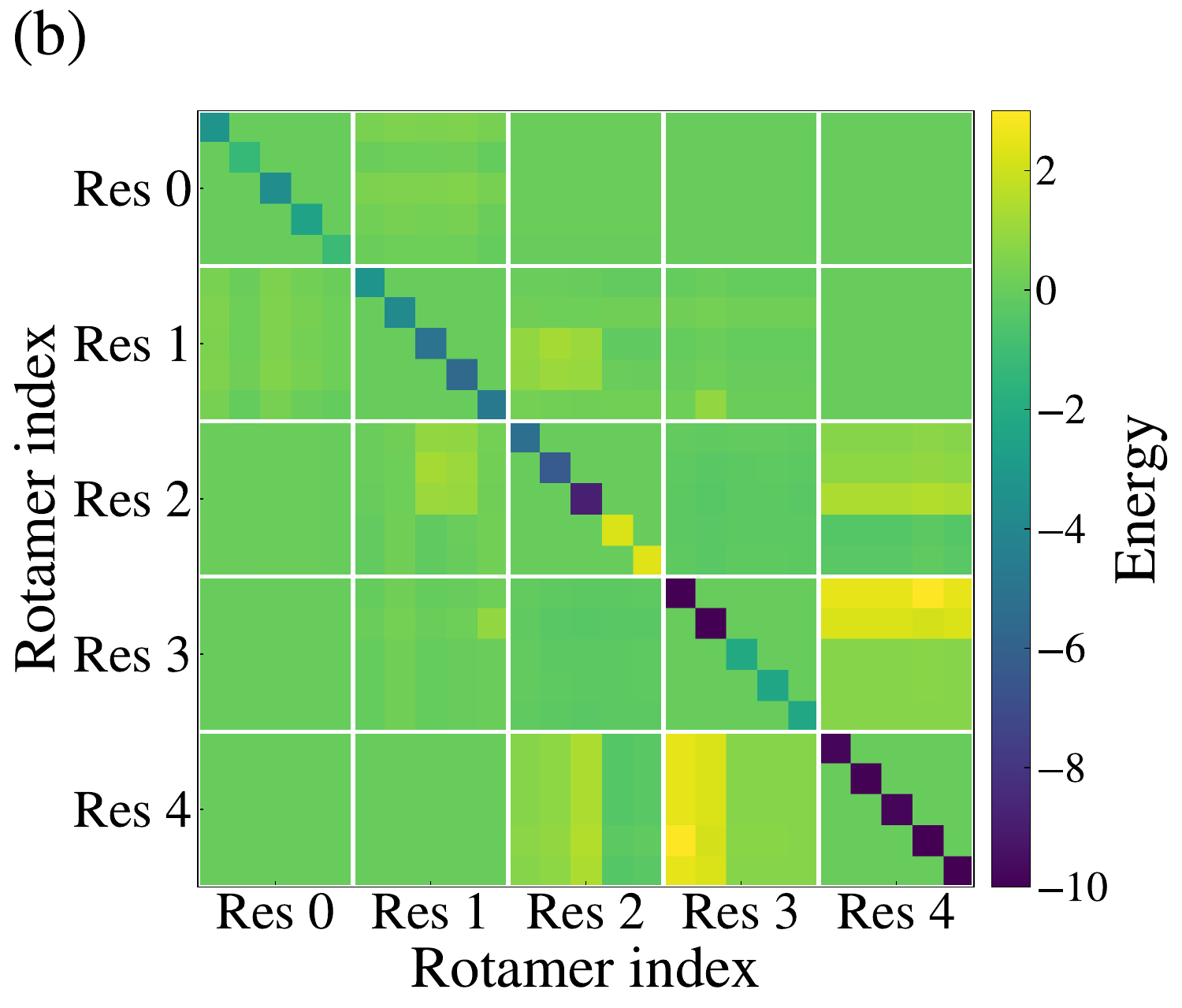}
    \caption{Correlation matrix of the characterising energies for (a) 4 residue, 4 rotamer structure and (b) 5 residue, 5 rotamer structure. The full Hamiltonian is represented, including all pairwise rotamer interactions. The Hamiltonians restricted to interactions between rotamers on nearest-neighbouring residues consists only of the residue-level tridiagonals of the full Hamiltonians.}
    \label{fig:full_vs_NN_combined}
\end{figure}

Fig. \ref{fig:full_vs_NN_combined} presents the correlation matrices of the interaction energies for two selected protein structures. Note here that these are energies from PyRosetta, which expresses energy in an arbitrary unit of measurement called REU (Rosetta Energy Unit). This unit is based on a combination of physics-based and statistics-based potentials, it does not match up with physical energy units. The matrices represent the full Hamiltonians, including all pairwise interactions between rotamers across residues. The nearest-neighbour-only Hamiltonians correspond to the three central diagonals (the tri-block-diagonal structure), which capture interactions restricted to rotamers on adjacent residues. Each matrix element denotes the interaction energy between a specific pair of rotamer configurations assigned to a residue pair.

The diagonal elements correspond to one-body energy terms, which dominate the overall energy landscape. The immediate off-diagonal elements capture interactions between rotamers assigned to nearest-neighbour residues—distinct from interactions between rotamers on the same residue—and constitute the next most significant contributions
The remaining off-diagonal elements reflect interactions between rotamers on non-neighbouring residues, which are largely negligible in magnitude. This observation supports our approximation of restricting the Hamiltonian to nearest-neighbour residue interactions in the current analysis. 

\refstepcounter{section}
\section*{Appendix \Alph{section}: \gls{MPS} \& \gls{SA} Convergence Rations}
\addcontentsline{toc}{section}{Appendix \Alph{section}: \gls{MPS} \& \gls{SA} Success Rates}
\label{app:mps_sa_success}
Results of the \gls{MPS-QAOA} simulations are averaged over several independent trajectories with different initial parameter vectors. Tables~\ref{table:MPS-success-res5} and~\ref{table:MPS-success-res6} show the total number of independent trajectories, and the convergence ratios for \gls{MPS-QAOA} simulations presented in Fig. \ref{fig:cpu_vs_qpu}.

Similarly, Tables~\ref{table:SA-success-res5} and~\ref{table:SA-success-res6} show the convergence ratios of the \gls{SA} simulations. All systems were simulated over 500 independent trajectories.

The $r^2$ values for the linear regression fits of the logarithmic data of the three methods are reported in Table \ref{table:r2}.

\begin{table}[H]
    \centering
    \small
    \setlength\tabcolsep{6pt}
    \renewcommand{\arraystretch}{1.2}

    \begin{tabular}{|c|c|c|c|}
        \hline
        Res. & Rot. & Total & Success Ratio \\
        \hline \hline
        5 & 5 & 1000 & 0.996 \\
        5 & 6 & 1000 & 0.996 \\
        5 & 7 & 1000 & 0.871 \\
        5 & 8 & 1000 & 0.781 \\
        5 & 9 & 1000 & 0.713 \\
        5 & 10 & 1000 & 0.496 \\
        \hline
    \end{tabular}
    \caption{Convergence ratios for \gls{MPS} simulations with 5 residues ($p=4$).}
    \label{table:MPS-success-res5}

    \vspace{1.2em}

    \begin{tabular}{|c|c|c|c|}
        \hline
        Res. & Rot. & Total & Success Ratio \\
        \hline \hline
        6 & 4 & 200 & 1.000 \\
        6 & 5 & 200 & 1.000 \\
        6 & 6 & 200 & 1.000 \\
        6 & 7 & 200 & 0.990 \\
        6 & 8 & 198 & 0.712 \\
        6 & 9 & 158 & 0.519 \\
        \hline
    \end{tabular}
    \caption{Convergence ratios for \gls{MPS} simulations with 6 residues ($p=25$).}
    \label{table:MPS-success-res6}

    \vspace{1.2em}

    \begin{tabular}{|c|c|c|}
        \hline
        Res. & Rot. & Success Ratio \\ 
        \hline \hline
        5 & 5 & 0.99 \\
        5 & 6 & 1.00 \\
        5 & 7 & 1.00 \\
        5 & 8 & 0.98 \\
        5 & 9 & 0.91 \\
        5 & 10 & 0.86 \\
        \hline
    \end{tabular}
    \caption{Success ratio for \gls{SA} simulations with 5 residues.}
    \label{table:SA-success-res5}

    \vspace{1.2em}

    \begin{tabular}{|c|c|c|}
        \hline
        Res. & Rot. & Success Ratio \\ 
        \hline \hline
        6 & 4 & 0.70 \\
        6 & 5 & 0.40 \\
        6 & 6 & 0.92 \\
        6 & 7 & 0.20 \\
        6 & 8 & 0.11 \\
        6 & 9 & 0.17 \\
        \hline
    \end{tabular}
    \caption{Success ratio for \gls{SA} simulations with 6 residues.}
    \label{table:SA-success-res6}

    \vspace{1.2em}

    \setlength\tabcolsep{10pt}
    \begin{tabular}{|c|c|}
        \hline
        Method & $r^2$ \\
        \hline \hline
        MPS QAOA 5 Res. & 0.978 \\
        MPS QAOA 6 Res. & 0.987 \\
        SA 5 Res. & 0.996 \\
        SA 6 Res. & 0.971 \\
        SV QAOA & 0.219 \\
        \hline
    \end{tabular}
    \caption{$r^2$ values for the fits of the results.}
    \label{table:r2}
\end{table}

\end{document}